\documentclass[prl,aps,twocolumn,superscriptaddress]{revtex4-1}

\usepackage{amsmath,amsfonts,amssymb,amsthm,mathtools}
\usepackage[locale=US]{siunitx}
\usepackage{graphicx}
\usepackage{color}
\usepackage{hyperref}
\usepackage{natbib}

\newcommand{\be}{\begin{equation}}
\newcommand{\ee}{\end{equation}}

\newcommand{\beq}{\begin{eqnarray}}
\newcommand{\eeq}{\end{eqnarray}}
\newcommand{\bra}[1]{\ensuremath{\langle#1|}}
\newcommand{\ket}[1]{\ensuremath{|#1\rangle}}

\newcommand{\mean}[1]{\ensuremath{\big\langle #1 \big\rangle}}

\begin{document}

\title{Heralded Generation of Macroscopic Superposition States \\ in a Spinor Bose-Einstein Condensate}

\author{L. Pezz\`{e}}
\affiliation{QSTAR, INO-CNR, and LENS, Largo Enrico Fermi 2, IT-50125 Firenze, Italy}
\author{M. Gessner}
\affiliation{QSTAR, INO-CNR, and LENS, Largo Enrico Fermi 2, IT-50125 Firenze, Italy}
\author{P. Feldmann}
\affiliation{Institut f\"ur Theoretische Physik, Leibniz Universit\"at Hannover, Appelstr. 2, DE-30167 Hannover, Germany}
\author{C. Klempt}
\affiliation{Institut f\"ur Quantenoptik, Leibniz Universit\"at Hannover, Welfengarten 1, DE-30167 Hannover, Germany}
\author{L. Santos}
\affiliation{Institut f\"ur Theoretische Physik, Leibniz Universit\"at Hannover, Appelstr. 2, DE-30167 Hannover, Germany}
\author{A. Smerzi}
\affiliation{QSTAR, INO-CNR, and LENS, Largo Enrico Fermi 2, IT-50125 Firenze, Italy}

\begin{abstract}
We propose to generate macroscopic superposition states of a large number of atoms 
in the ground state of a three-mode spinor Bose-Einstein condensate. 
The ground state is protected by a finite energy gap, is immune to phase noise,  and the measurement of the number 
of particles in one mode heralds the coherent preparation of the two 
other modes simultaneously empty and filled. 
Highly entangled macroscopic superposition states are generated with large probability also taking into account 
quasi-adiabatic preparation of the ground state.  
All the key ingredients necessary to realize our proposal are experimentally available.
\end{abstract}

\maketitle
\date{\today}

{\it Introduction.---}Macroscopic superposition states (MSSs)
are a striking prediction of quantum theory~\cite{Schrodinger} as well as
a holy grail in quantum science and technology.
Prominent examples, such as GHZ states of many qubits or NOON states of bosonic particles,
set a benchmark in quantum metrology~\cite{PezzeRMP} 
and can find important application in quantum information processing~\cite{PanRMP}. 
Furthermore, they strongly couple to the environment and may thus provide a fundamental testbed of decoherence processes  
unveiling the transition from quantum to classical~\cite{Zurek}. 
However, it is extremely challenging to generate genuine MSSs, {\it i.e.}~for a large number of particles.

For photons, there exist several proposals~\cite{PezzePRL2008, CablePRL2007, MatthewsPRL2011} for the creation of
NOON states $(\ket{N}_a \ket{0}_b + \ket{0}_a \ket{N}_b)/\sqrt{2}$, where $a$ and $b$ are two polarization or spatial modes
simultaneously empty and filled with $N$ particles.
So far, these states have been created up to $N=5$ photons~\cite{AfekSCIENCE2010, MitchellNATURE2004, WaltherNATURE2004}
-- for instance mixing a coherent and a squeezed-vacuum state at a balanced beam splitter and post-selecting the 
total number of photons in output~\cite{AfekSCIENCE2010, PezzePRL2008} --
and $N=10$ indistinguishable nuclear spins~\cite{JonesSCIENCE2009}.
Maximally entangled GHZ states $(\ket{\uparrow}^{\otimes N} + \ket{\downarrow}^{\otimes N})/\sqrt{2}$
of $N$ spin-1/2 have been generated up to $N=8$ photons~\cite{YaoNATPHOT2012} combining 
several parametric down-conversion sources,
and up to $N =14$ trapped ions using one-axis 
twisting dynamics~\cite{LeibfriedNATURE2005, MonzPRL2011, KitagawaPRA1993, MolmerPRL1999}. 
Such a nonlinear evolution~\cite{KitagawaPRA1993} has also been realized in systems of a large number of particles~\cite{PezzeRMP}
using atom-atom interaction in Bose-Einstein condensates (BECs) or cavity-assisted light-matter interaction in cold-thermal atom ensembles.  
Yet, the generation of MSSs with this method~\cite{MicheliPRA2003, PezzePRL2009, PawlowskiPRA2017} 
requires long evolution times and, so far,  
spin-squeezed~\cite{LerouxPRL2010, GrossNATURE2010, RiedelNATURE2010, BohnetSCIENCE2016} 
and slightly non-Gaussian states~\cite{StrobelSCIENCE2014} have been successfully identified. 
MSSs can be also generated adiabatically, for instance preparing the 
ground state of a BEC trapped in a double-well potential with attractive interaction~\cite{CiracPRA1998, TrenkwalderNATPHYS2016}. 
However, due to the degeneracy of the spectrum, this method is extremely fragile to symmetry-breaking perturbations.  

Here we demonstrate that MSSs of a large number of atoms can be created in the ground state of a ferromagnetic spin-1 BEC. 
After the preparation of the ground state, MSSs are generated stochastically 
with large (up to 90\%) probability and heralded by the measurement 
of the number of atoms in one of the three modes of the BEC.
The state creation is nondestructive -- the MSS can be manipulated after its generation -- and 
scalable with the number of particles.
Remarkably, the ground state is protected by a finite energy gap and lives in a decoherence-free subspace robust against phase noise. 
Furthermore, even without any post-selection, the ground state of the spinor BEC has a Fisher information 
scaling at the Heisenberg limit and, following~\cite{KrusePRL2016}, can thus find 
key applications in quantum sensing~\cite{GabbrielliPRL2015, WuPRA2016, SzigetiPRL2017, KajtochPREPRINT}.
The state preparation requires crossing over a quantum phase transition point~\cite{ZhangPRL2013}, as  
recently demonstrated experimentally~\cite{HoangPNAS2016, LuoSCIENCE2017} --
note however that the states discussed here have not been addressed in these experiments.  
We thus believe, supported by numerical simulations, that our method can provide a feasible 
scheme to generate MSSs of a large number of particles in current experiments. 

{\it The model.---}A spin-$1$ BEC of $N$ particles with magnetic 
sublevels $m_F=\{0,\pm 1\}$ is described in the single-mode approximation~\cite{LawPRL1998, Stamper-KurnRMP2013} by the Hamiltonian
(energy is in units of $|\lambda|$)
\begin{align}
\begin{split} \label{eq:Hamiltonian}
\hat{H} =& \left ( \frac{q}{ \vert \lambda \vert} + \frac{1}{2} - \hat{N}_0 \right)(\hat{N}_{+1} + \hat{N}_{-1}) 
 - (\hat{a}_1^\dagger \hat{a}_{-1}^\dagger \hat{a}_0^2 + \operatorname{h.\!c.}),
\end{split}
\end{align}
where $\hat{a}_{m_F}^\dagger$ and $\hat{a}_{m_F}$ are bosonic creation and annihilation operators, respectively, 
and $\hat{N}_{m_F} = \hat{a}^\dagger_{m_F} \hat{a}_{m_F}$ are the associated number operators. 
The interaction coefficient $\lambda$ depends on the scattering lengths, atom mass, and trapping potential~\cite{LawPRL1998, nota}. 
In the following we consider ferromagnetic interaction $\lambda<0$ as in $^{87}$Rb atoms~\cite{Stamper-KurnRMP2013}. We have
$q = (\Delta E_{1}+ \Delta E_{-1})/2$, where $\Delta E_{m_F} = E_{m_F} - E_0$ is the 
relative energy shift of the $m_F= \pm 1$ mode with respect to $m_F=0$, which can be tuned by an external magnetic field 
or near-resonant microwave dressing~\cite{Stamper-KurnRMP2013}.
Spin-changing collisions, described by the last part of Eq.~\eqref{eq:Hamiltonian}, 
preserve the total magnetization $\hat{D}\equiv \hat{N}_{+1}-\hat{N}_{-1}$.
The Hilbert space of states with $\hat{D}|\psi\rangle=0$ is spanned by the Fock state basis  
$|k\rangle\equiv \ket{k}_{+1} \ket{N-2k}_0 \ket{k}_{+1}$ with 
$N = N_{-1} + N_0 + N_{+1}$ the total, fixed, number of particles in the three modes. 
To have a better insight to Eq.~(\ref{eq:Hamiltonian}), let us 
introduce symmetric ($g$) and antisymmetric ($h$) combinations of operators $\hat{a}^{\dag}_{\pm 1}$:
$\hat{g}^\dagger=(\hat{a}_1^\dagger+\hat{a}_{-1}^\dagger)/\sqrt{2}$ 
and $\hat{h}^\dagger=(\hat{a}_1^\dagger-\hat{a}_{-1}^\dagger)/\sqrt{2}$.
We can form two non-commuting sets of collective pseudospin-$\frac{1}{2}$ 
operators:
\begin{equation}
\begin{alignedat}{2} \label{SAoperators}
\hat{S}_x &= \frac{\hat{a}_0^\dag\hat{g} + \hat{g}^\dag\hat{a}_0}{2},
\hspace{3em}
\hat{A}_x &= \frac{\hat{a}_0^\dag\hat{h} + \hat{h}^\dag\hat{a}_0}{2},\\
\hat{S}_y &= \frac{\hat{a}_0^\dag\hat{g} - \hat{g}^\dag\hat{a}_0}{2i},
\hspace{3em}
\hat{A}_y &= \frac{\hat{a}_0^\dag\hat{h} - \hat{h}^\dag\hat{a}_0}{2i}, \\
\hat{S}_z &= \frac{\hat{a}_0^\dag\hat{a}_0 -\hat{g}^\dag\hat{g}}{2},
\hspace{3em}
\hat{A}_z &= \frac{\hat{a}_0^\dag\hat{a}_0 -\hat{h}^\dag\hat{h}}{2}.
\end{alignedat}
\end{equation}
Thus $\hat{\mathbf{S}}\equiv (\hat{S}_x,\hat S_y,\hat S_z)$ describes the two-mode system composed by modes 
$\hat{a}_0$ and $\hat{g}$, and
$\hat{\mathbf{A}}\equiv (\hat A_x,\hat A_y,\hat A_z)$ corresponds to modes $\hat{a}_0$ and $\hat{h}$.
We will also consider the two-mode system composed by modes $\hat{a}_{+1}$ and $\hat{a}_{-1}$, 
$\hat{\mathbf{J}}\equiv (\hat{J}_x,\hat J_y,\hat J_z)$, where 
$\hat{J}_x = (\hat{a}_1^\dag\hat{a}_{-1} + \hat{a}_{-1}^\dag\hat{a}_1)/2$, 
$\hat{J}_y = (\hat{a}_1^\dag\hat{a}_{-1} - \hat{a}_{-1}^\dag\hat{a}_1)/2i$ and 
$\hat{J}_z = (\hat{a}_1^\dag\hat{a}_1 -\hat{a}_{-1}^\dag\hat{a}_{-1})/2$.
Equation~(\ref{eq:Hamiltonian}) rewrites as
\begin{equation} \label{eq.Hamiltonian.SA}
\hat H = -2 \left( \hat S_x^2 + \frac{q}{3 |\lambda|} \hat S_z\right) - 2 \left(\hat A_y^2 + \frac{q}{3 |\lambda|}\hat A_z\right),
\end{equation}
up to constant terms~\cite{PezzeRMP}.

\begin{figure}[t]
 \begin{center}
 \includegraphics[width=\columnwidth]{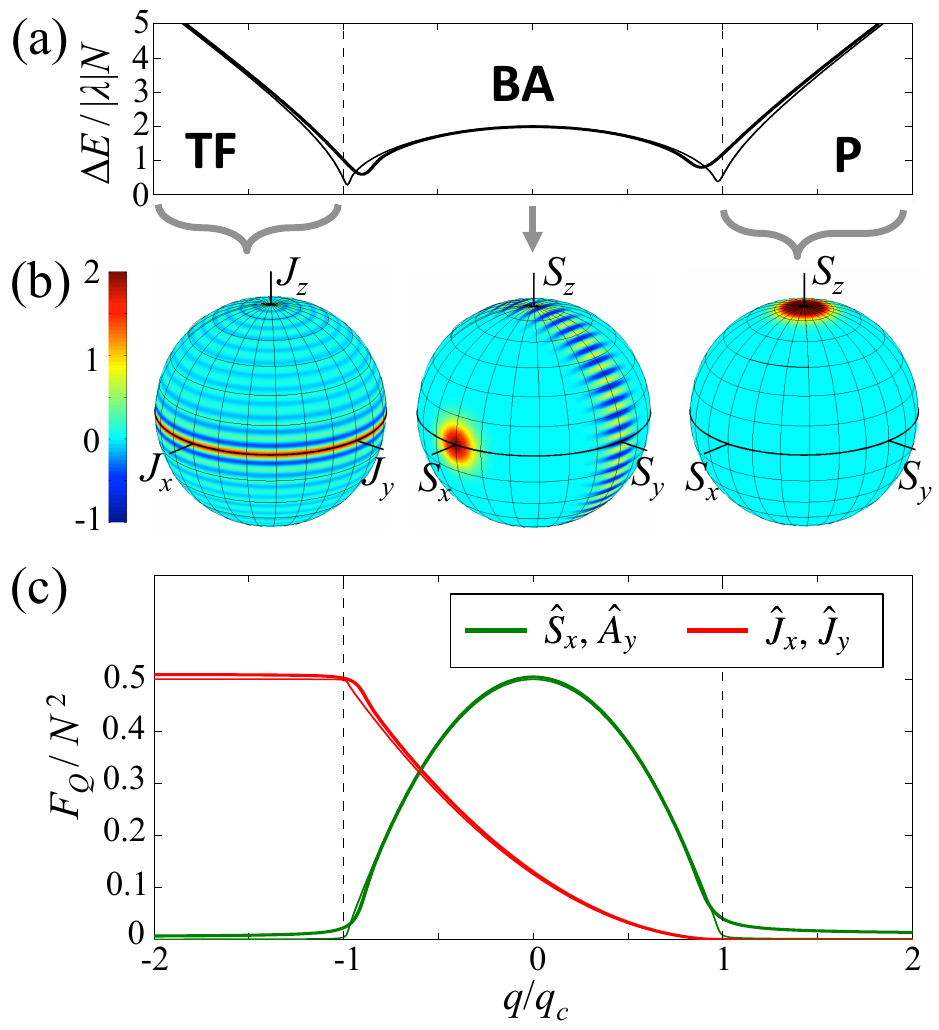}
\end{center}
\caption{Ground state of a ferromagnetic spinor BEC.
(a) Energy gap  $\Delta E/(|\lambda|N)$ between the ground state and the first excited state as a function of $q/q_c$.
(b) Wigner distribution of the ground state in the Bloch sphere. 
In the P phase (right), the two mode distribution is obtained after  tracing out the $h$ mode and is
given by a coherent spin state polarized in on the $S_z$ axis. 
In the TF phase (left) tracing out the $m_F=0$ mode gives a twin Fock state on the $J_x$-$J_y$ equator.
For $q=0$ (central sphere) we plot the Wigner distribution of the heralded state obtained after the measurement of 
number of particles in the $h$ mode (see text): with the largest probability this provides a NOON-like 
state along the $S_x$ axis.
(c) QFI of the (full) ground state for different operators. 
In all panels (a) and (c), thick likes are for $N=100$ and thin lines for $N=1000$.
The quantum phase transition points are indicated by dashed vertical lines.
$N=100$ in panel (b).
}
\label{fig1}
\end{figure}

The Hamiltonian $\hat{H}$ has, as a function of the ratio $q/|\lambda|$, 
three quantum phases~\cite{ZhangPRL2013, HoangPNAS2016, LuoSCIENCE2017} in the thermodynamic limit 
($N \to \infty$, $|\lambda|\to 0$ and $N |\lambda|={\rm const.}$). 
The quantum phase transition points $q=\pm q_c$, where $q_c = 2 N |\lambda|$, are identified by an 
energy gap that closes as $N^{-1/3}$~\cite{ZhangPRL2013}, see Fig.~\ref{fig1}(a).
For $q>q_c$, we have a polar~(P) phase, where the ground state has all particles in $m_f=0$, 
$|\psi_{\rm gs}^{q>q_c} \rangle = \ket{0}_{-1}\ket{N}_0 \ket{0}_{+1}$.
Tracing out the $h$ (or the $g$) mode provides $\ket{N}_0 \ket{0}_{g}$ ($\ket{N}_0 \ket{0}_{h}$)
that corresponds to a coherent spin state polarized along the $S_z$ ($A_z$) axis, see Fig.~\ref{fig1}(b).
For $|q|<q_c$, we have a broken-axisymmetry~(BA) phase, where all three modes are populated, 
with $\mean{\hat{N}_0} = \tfrac{N}{2} ( 1 + \frac{q}{q_c})$~\cite{PezzeRMP}.
In particular, at $q=0$~\cite{nota1}, the Hamiltonian~(\ref{eq.Hamiltonian.SA}) takes the form 
$\hat H = - 2 \hat S_x^2 -2  \hat A_y^2$. 
MSSs can be heralded by measuring the number of particles in the $h$ (or in the $g$) mode. 
A simple intuition can be gained by considering the case in which the measurement of particles in $h$ yields $N_h=0$. 
In this case, we may argue (see a detailed analysis below) that the contribution of $\hat{A}_y$ to $\hat{H}$ can be disregarded. 
The ground state of $\hat{H} \approx - 2 \hat{S}_x^2$ is a NOON state along the $S_x$ axis, see Fig.~\ref{fig1}(b).:
$\exp(-i\tfrac{\pi}{2}\hat S_y)(\ket{N}_g \ket{0}_0 +  \ket{0}_g \ket{N}_0) /\sqrt{2}$, corresponding
(up to a rotation) to a superposition of $N$ atoms in the symmetric $g$ mode and $0$ atoms in the $m_f=0$ mode, and viceversa. 
As shown below, the result $N_h=0$ has the highest probability.
States heralded by other measurement results will be analyzed in the following.
For $q<-q_c$, we identify a Twin-Fock (TF) phase, where the condensate in $m_f=0$ is completely depleted and 
the ground state is given by $\ket{\psi_{\rm gs}^{q<q_c}} =  \ket{N/2}_{-1} \ket{0}_0 \ket{N/2}_{+1}$ 
with exactly $N/2$ particles in $m_f=\pm 1$~\cite{ZhangPRL2013, LuoSCIENCE2017}.
Tracing out the $m_f=0$ modes provides $\ket{N/2}_{-1} \ket{N/2}_{+1}$ that corresponds 
to a Twin-Fock state~\cite{HollandPRL1993}, see Fig.~\ref{fig1}(b).

\begin{figure*}
 \begin{center}
 \includegraphics[width=\textwidth]{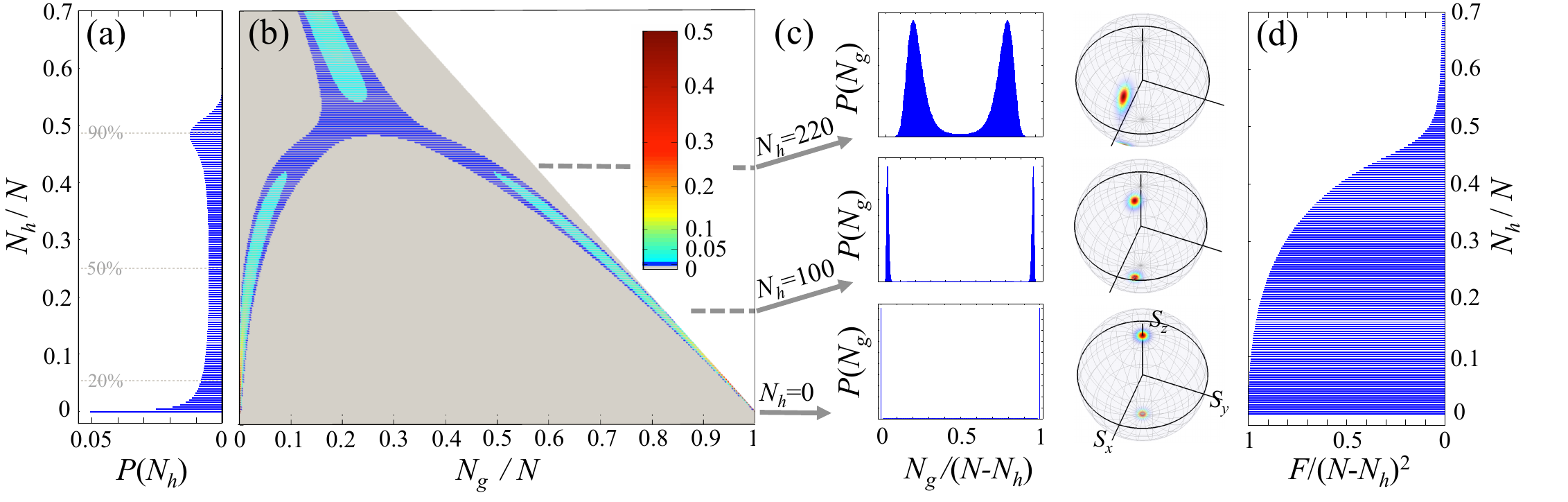}
\end{center}
\caption{Stochastic generation of MSSs. 
We consider the ground state of Eq.~(\ref{eq:Hamiltonian}) at $q=0$ and measure the number of particles in the $h$ mode.  
This measurement heralds the preparation of a state $\ket{\phi_{N_h}}$ in the $g-0$ modes.
(a) Probability to detect $N_h$ particles (bars) in the ground state. 
The dashed lines are cumulative probability thresholds.
(b) Heralded states $\ket{\phi_{N_h}}$.
To help the visualization, the heralded states have been rotated by $\exp(-i \tfrac{\pi}{2} \hat{S}_y)$.
We show the probability distribution of $N_g$ particles in the $g$ mode obtained after 
measuring $N_h$ particles in the $h$ mode, $P(N_g) = \vert \bra{N_g, N-N_g} e^{-i (\pi/2) \hat{S}_y} \ket{\phi_{N_h} } \vert^2$.
Notice that $N_g$ ranges from $0$ to $N-N_h$.
The two branches of the distribution for $N_h \lesssim N/2$ are due to the presence of MSSs.
(c) Examples of probability distribution $P(N_g)$ (left) and Husimi 
distribution in the $\mathbf{S}$ manifold (right) of the rotated heralded state, 
shown here for certain values of $N_h$: $N_h=0$ (bottom panels), $N_h=100$ (middle) and $N_h=220$ (top).
(d) QFI $F_Q[\ket{\phi_{N_h}}, \hat{S}_x]$ of the state obtained in the $g-0$ modes after projection, as a function of $N_h$.
In all panels $N=500$.
}
\label{fig2}
\end{figure*}

To further characterize the ground state, we determine its interferometric sensitivity as well as its multiparticle entanglement. 
Both quantities are captured by the quantum Fisher information (QFI) $F_Q[\ket{\psi}, \hat{R}]$, which characterizes the sensitivity of the state $\ket{\psi}$ under unitary rotations generated by $\hat{R}$ \cite{PezzeVarenna}. 
A plot of the QFI is shown in Fig.~\ref{fig1}(c). 
At $q=0$ we have 
\begin{equation} \label{FQCBA}
F_Q\big[\ket{\psi_{\rm gs}^{q=0}}, \hat{R}_\mathrm{opt} \big] = N(N+1)/2, 
\end{equation}
where the optimal rotation operator, $\hat{R}_\mathrm{opt}$, is given by an arbitrary linear combination of 
$\hat{S}_x$ and $\hat{A}_y$. 
In the TF phase, 
\begin{equation} \label{FQTF}
F_Q\big[\ket{\psi_{\rm gs}^{q<-q_c}}, \hat{R}_\mathrm{opt} \big]=N(N+2)/2,
\end{equation}
where $\hat{R}_\mathrm{opt}$ is an arbitrary linear combination of 
$\hat{J}_x$ and $\hat{J}_y$.
In both cases, only two of the three modes are involved in the optimal rotation and 
$F_Q[\ket{\psi}, \hat{R}_\mathrm{opt}]/N$ scales as the number of entangled particles~\cite{PezzeRMP, HyllusPRA2012}.
Hence, we observe that, in addition to the Twin-Fock state, the ground state at $q = 0$ 
possesses large entanglement useful for metrology.
This can be exploited with coherent manipulations of the state on the $\mathbf{\hat{S}}$-Bloch sphere
using rf pulses~\cite{PeiseNATCOMM2015, GrossNATURE2011} 
as has been recently demonstrated with an atomic clock~\cite{KrusePRL2016}.

{\it Heralded generation of MSSs.--}
In the following we focus on the ground state at $q=0$ \cite{nota1} and analyze (without any approximation) the state $\ket{\phi_{N_h}}$ 
of the $g$-$0$ modes obtained after measuring $N_h$ particles in the $h$ mode.
The probability to measure $N_h$ particles in mode $h$ is shown in Fig.~\ref{fig2}(a). 
The highest probability is found for $N_h=0$, as discussed above, and only even values of $N_h$ are detectable. 
The state heralded by the measurement result $N_h=0$ is (with a fidelity of more that 99\%) a rotated NOON state in the $g$-$0$ modes, 
$\ket{\phi_{N_h=0}} \approx \exp(-i\tfrac{\pi}{2}\hat S_y)(\ket{N}_g \ket{0}_0 +  \ket{0}_g \ket{N}_0) /\sqrt{2}$, see Fig.~\ref{fig2}(b)~and~(c). 
Moreover, MSSs (given by a coherent superposition of highly populated $0$ and $g$ modes) 
are heralded not only for $N_h=0$, but also for values of $0\leq N_h \lesssim N/2$, see Fig.~\ref{fig2}(b).
The population of the $m_F=0$ and $g$ modes start to overlap around $N_h \approx N/2$.
Remarkably, as the probability to detect $N_h \leq N/2$ is about 90\%, we can conclude that the measurement of $N_h$
prepares a MSS in the $g$-$0$ modes in 90\% of the cases, regardless of the total number of atoms. 
To emphasize the coherence of the states $\ket{\phi_{N_h}}$ obtained after heralding projection, we show, in Fig.~\ref{fig2}(c), their QFI.
Values $F_Q[\ket{\phi_{N_h}}, \hat{S}_x] \sim (N-N_h)^2$ are found up to $N_h \approx N/2$~\cite{nota2}. 

\begin{figure}
 \begin{center}
 \includegraphics[width=\columnwidth]{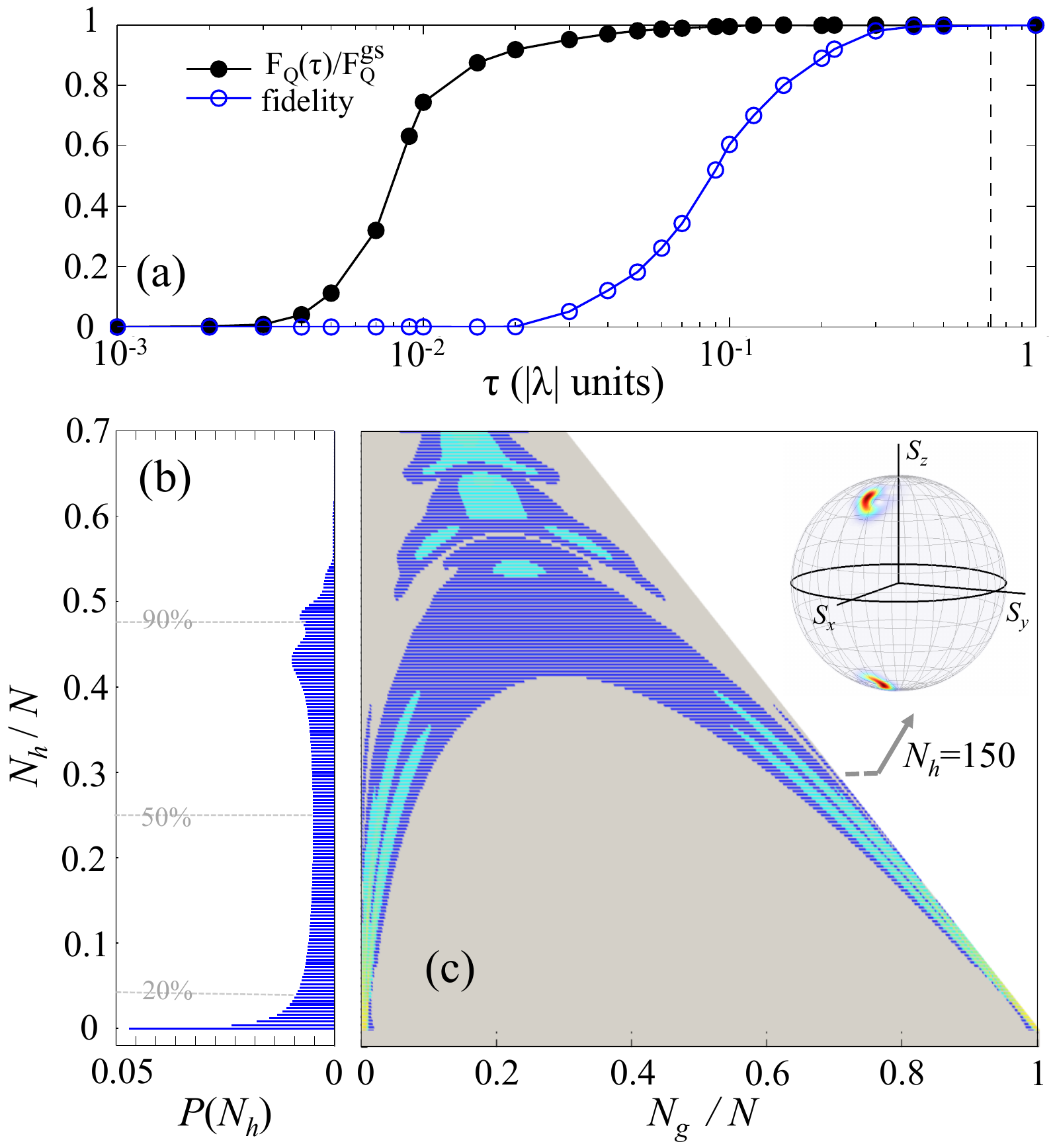}
\end{center}
\caption{Stochastic generation of MSSs via quasi-adiabatic evolution.
We consider the state $\ket{\psi(\tau)}$ prepared at $q=0$ in a ramping time $\tau$ starting from the P phase.  
(a) Fidelity with the ground state at $q=0$, $|\bra{\psi_{\rm gs}} \psi(\tau) \rangle|^2$ (blue circles), and 
ratio between the QFI of the evolved state and the one for the ground state, $F_Q(\tau)/F_Q^{\rm gs}$ (black dots), 
with $F_Q(\tau) = F_Q[\ket{\psi(\tau)}, \hat{S}_x]$
and 
$F_Q^{\rm gs} = N(N+1)/2$, Eq.~(\ref{FQCBA}), as a function of $\tau$.
The dashed line is the inverse of the energy gap at $q=q_c$.
Solid lines are a guide to the eye.
(b) Probability to detect $N_h$ particles (bars) for the state $\ket{\psi(\tau)}$ prepared at $\tau = 0.06$.
The dashed lines are cumulative probability thresholds.
(c) Heralded state after a rotation $\exp(-i \tfrac{\pi}{2} \hat{S}_y)$.
We show the probability distribution of $N_g$ particles in the $g$ mode obtained after measuring $N_h$ particles in the $h$ mode.
The inset shows the Husimi distribution of the heralded state at $N_h=150$, as an example.
In all panels $N=500$.}
\label{fig3}
\end{figure}

{\it Robustness and experimental feasibility.--}We propose to address the stochastic generation of MSSs by 
(i) initially preparing a condensate in the P phase, at $q > q_c$, (ii) slowly decreasing $q$ down to $q=0$
and (iii) measuring the number of particles in the $h$ mode.
Preparing the polarized state allows us to address the subspace of zero magnetization 
($\hat{D} \ket{\psi} = 0$), which is conserved during the adiabatic process.
This entails stability with respect to phase noise generated by $\hat D$: 
for any obtained state $|\psi\rangle$ and any time-dependent noise parameter 
$\phi(t)$, $\exp(-i\phi(t)\hat D)|\psi\rangle=|\psi\rangle$. 
The linear coupling to the magnetic field and its fluctuations (linear Zeeman shift) become irrelevant.
Furthermore, after passing though the quantum phase transition point at $q = q_c$, the energy gap re-opens, see Fig~\ref{fig1}(a). 
Crucially, at $q = 0$ the energy gap between the ground state and the first excited
state assumes its largest value in the BA phase.
This gap protects $\ket{\psi_{\rm gs}^{q=0}}$ against perturbations, {\it e.g.} due to finite temperature.  

A quasi-adiabatic state preparation similar to the one proposed here 
has been demonstrated experimentally in~\cite{HoangPNAS2016, LuoSCIENCE2017}
with the aim of addressing the TF phase~\cite{LuoSCIENCE2017, ZhangPRL2013}, 
which requires decreasing $q$ beyond $q =0$ and crossing the second quantum phase transition point at $q=-q_c$. Preparing the state at $q=0$ may be even less demanding in terms of BEC stability.

We have numerically simulated the quasi-adiabatic state preparation of the $q = 0$ ground state with realistic parameters:
we assume a BEC prepared at $t=0$ at $q/q_c=2$ in the P phase, and decrease $q$
linearly as $q(t) / q_c = 2- t / \tau$, where the ramping time $\tau$ characterizes the non-adiabaticity of the process.
In Fig.~\ref{fig3}(a) we plot the overlap (blue dots) between the evolved state $\ket{\psi(\tau)}$ 
and the ground state $\ket{\psi_{\rm gs}}$ at $q=0$.
We also show the ratio between the QFI of the evolved $\ket{\psi(\tau)}$ and the full adiabatic value, Eq.~(\ref{FQCBA}).
These results are remarkable: even for a relatively fast ramping, faster than the inverse critical gap and thus
producing small values of the fidelity $|\langle \psi_{\rm gs} \ket{\psi(\tau)}|^2$, the
QFI is hardly affected by the finite ramping speed and we obtain quite high values, comparable with the one of the adiabatic ground state.
In Fig.~\ref{fig3}(b) and (c) we show results for the heralded state generation for a ramping time $\tau = 0.06$, 
that is feasible experimentally~\cite{LuoSCIENCE2017, HoangPNAS2016} taking into account that 
$\vert \lambda \vert \sim 1 \div 10$ mHz and ramping times can be of the order of $ \sim 10$ seconds. 
Figure~\ref{fig3}(b) shows the distribution of particles in the $h$ mode, while  
Fig.~\ref{fig3}(c) shows the heralded states (in terms of distribution of particles in the $g$ mode after 
measurement of $N_h$ particles in the $h$ mode), as in Fig.~\ref{fig2}(b). 
MSSs are found about 80\% of the cases
(a value slightly lower than in the full adiabatic case).
While, as expected, the finite ramping time broadens the particle distribution, 
the effect is not dramatic and MSSs 
are clearly reachable for the experimental ramping time and for a wide range of $N_h$. 
Finally, we point out that the $h$ and $g$ modes can be addressed after a $\pi/2$ balanced beam splitter between the $m_f = \pm 1$
with a two-frequency microwave pulse, as demonstrated in Ref.~\cite{LuckeSCIENCE2011}.
The measurement of number of particles in the $h$ mode is thus obtained after a $\pi/2$ coupling between 
$m_f = \pm 1$ (i.e. applying $\exp[-i (\pi/2) \hat{J}_x]$)
and the measurement of number of particles in one of the $m_f = \pm 1$ modes.
We also emphasize that the state preparation is symmetric in the $h$ and $g$ modes: the measurement of number of particles
in the $g$ mode would prepare MSSs in the $h$ mode with the same distributions discussed above. 
Finally, the generation of MSSs does not rely on single-particle resolution in the heralding detection, 
as is the case in many protocols: a small, finite, detection noise does not remove the macroscopic superposition features.

{\it Conclusions.---}We have discussed a method to create macroscopic superposition states 
of a large number of atoms in a spinor Bose-Einstein condensate. 
These states are prepared stochastically with high probability and are heralded by the measurement of the 
number of particles in one of the modes (namely the symmetric or the anti-symmetric combination of hyperfine 
$\pm 1$ states).
Moreover, even without post-selection, the state generated by quasi-adiabatic evolution is highly entangled and has a 
quantum Fisher information showing Heisenberg scaling with the number of atoms.
The crucial advantage of our method is the robustness with respect to experimental phase noise 
and the presence of a finite energy gap which protects the highly entangled state. 
Finally, we point out that all ingredients necessary for the state generation discussed in this manuscript are 
experimentally available following 
Refs.~\cite{LuoSCIENCE2017, HoangPNAS2016, KrusePRL2016, PeiseNATCOMM2015, LuckeSCIENCE2011, GrossNATURE2011}.
Our proposal thus provides a crucial step forward for the generation of  
macroscopic superposition states of many atoms, a highly sought but still missing 
ingredient of quantum state engineering for quantum technologies. 

\begin{acknowledgments}
We acknowledge support by the SFB 1227 ``DQ-mat'' of the German Research Foundation (DFG). 
M.\,G. thanks the Alexander von Humboldt foundation for support.
\end{acknowledgments}

\bibliographystyle{prsty}

\begin{thebibliography}{}

\bibitem{Schrodinger}
E. Schr\"odinger, 
Die gegenw\"artige Situation in der Quantenmechanik, 
Naturwissenschaften {\bf 23}, 807 (1935). 
For a translation in english see
Proc. of the Am. Phil. Soc. {\bf 124}, 323 (1980).

\bibitem{PezzeRMP}
L. Pezz\`e, A. Smerzi, M. K. Oberthaler, R. Schmied and P. T. Treutlein,
Quantum metrology with nonclassical states of atomic ensembles, 
arXiv:01609.1609.

\bibitem{PanRMP}
J.-W. Pan, Z.-B. Chen, C.-Y. Lu, H. Weinfurter, A. Zeilinger, and M. Zukowski,
Multiphoton entanglement and interferometry, 
Rev. Mod. Phys. {\bf 84}, 777 (2012).

\bibitem{Zurek}
W. H. Zurek, Decoherence and the transition from quantum to classical, 
Phys. Today {\bf 44}, 36 (1991); 
W. H. Zurek, Decoherence, einselection, and the quantum origins of the classical, 
Rev. Mod. Phys. {\bf 75}, 715 (2003).

\bibitem{PezzePRL2008}
L. Pezz\`e and A. Smerzi, 
Mach-Zehnder Interferometry at the Heisenberg Limit with Coherent and Squeezed-Vacuum Light, 
Phys. Rev. Lett. {\bf 100}, 073601 (2008);
H. F. Hofmann and T. Ono, 
High-photon-number path entanglement in the interference of spontaneously down-converted photon pairs with coherent laser light, 
Phys. Rev. A {\bf 76}, 031806(R) (2007).

\bibitem{CablePRL2007}
H. Cable and J. P. Dowling, 
Efficient Generation of Large Number-Path Entanglement Using Only Linear Optics and Feed-Forward,
Phys. Rev. Lett. {\bf 99}, 163604 (2007);
J. P. Dowling, 
Quantum optical metrology -- the lowdown on high-N00N states, 
Cont. Phys. {\bf 49}, 125 (2008).

\bibitem{MatthewsPRL2011}
J. C. F. Matthews, A. Politi, D. Bonneau, and J. L. O'Brien, 
Heralding Two-Photon and Four-Photon Path Entanglement on a Chip, 
Phys. Rev. Lett. {\bf 107}, 163602 (2011).

\bibitem{AfekSCIENCE2010}
I. Afek, O. Ambar, Y. Silberberg, 
High-NOON States by Mixing Quantum and Classical Light, 
Science {\bf 328}, 879 (2010).

\bibitem{MitchellNATURE2004}
M. W. Mitchell, J. S. Lundeen and A. M. Steinberg, 
Super-resolving phase measurements with a multiphoton entangled state,
Nature {\bf 429}, 161 (2004).

\bibitem{WaltherNATURE2004}
P. Walther, J.-W. Pan, M. Aspelmeyer, R. Ursin, S. Gasparoni and A. Zeilinger,
De Broglie wavelength of a non-local four-photon state,
Nature {\bf 429}, 158 (2004).

\bibitem{JonesSCIENCE2009}
J. A. Jones, S. D. Karlen, J. Fitzsimons, A. Ardavan, S. C. Benjamin, G. A. D. Briggs, J. J. L. Morton, 
Magnetic Field Sensing Beyond the Standard Quantum Limit Using 10-Spin NOON States, 
Science {\bf 342}, 1166 (2009).

\bibitem{YaoNATPHOT2012}
X.-C. Yao, T.-X. Wang, P. Xu, H. Lu, G.-S. Pan, X.-H. Bao, C.-Z. Peng, C.-Y. Lu, Y.-A. Chen and J.-W. Pan,
Observation of eight-photon entanglement,
Nat. Phot. {\bf 6}, 225 (2012).

\bibitem{LeibfriedNATURE2005}
D. Leibfried, M. D. Barrett, T. Schaetz, J. Britton, J. Chiaverini, W. M. Itano, J. D. Jost, C. Langer, and D. J. Wineland, 
Toward Heisenberg-limited spectroscopy with multiparticle entangled states, 
Science {\bf 304}, 1476 (2004);
D. Leibfried, E. Knill, S. Seidelin, J. Britton, R. B. Blakestad, J. Chiaverini, D. B. Hume, W. M. Itano, 
J. D. Jost, C. Langer, R. Ozeri, R. Reichle, and D. J. Wineland, 
Creation of a six-atom `Schr\"odinger cat' state, 
Nature {\bf 438}, 639 (2005).

\bibitem{MonzPRL2011}
T. Monz, P. Schindler, J. T. Barreiro, M. Chwalla, D. Nigg, W. A. Coish, M. Harlander, W. H\"ansel, M. Hennrich, and R. Blatt,
14-qubit entanglement: Creation and coherence, 
Phys. Rev. Lett. {\bf 106}, 130506 (2011).

\bibitem{MolmerPRL1999}
K. M\o lmer, and A. S\o rensen, 
Multiparticle entanglement of hot trapped ions, 
Phys. Rev. Lett. {\bf 82}, 1835 (1999)

\bibitem{KitagawaPRA1993}
M. Kitagawa, and M. Ueda, 
Squeezed spin states,
Phys. Rev. A {\bf 47}, 5138 (1993).

\bibitem{MicheliPRA2003}
A. Micheli, D. Jaksch, J. I. Cirac, and P. Zoller, 
Many-particle entanglement in two-component Bose-Einstein condensates, 
Phys. Rev. A {\bf 67}, 013607 (2003).

\bibitem{PezzePRL2009}
L. Pezz\`e and A. Smerzi, 
Entanglement, nonlinear dynamics, and the Heisenberg limit, 
Phys. Rev. Lett. {\bf 102}, 100401 (2009).

\bibitem{PawlowskiPRA2017}
F. Piazza, L. Pezz\`e, and A. Smerzi, 
Macroscopic superpositions of phase states with Bose-Einstein condensates, 
Phys. Rev. A {\bf 78}, 051601(R) (2008);
K. Pawlowski, M. Fadel and P. Treutlein, Y. Castin, and A. Sinatra,
Mesoscopic quantum superpositions in bimodal Bose-Einstein condensates:
Decoherence and strategies to counteract it, 
Phys. Rev. A {\bf 95}, 063609 (2017).

\bibitem{LerouxPRL2010}
I. D. Leroux, M. H. Schleier-Smith, and V. Vuleti\'c, 
Implementation of cavity squeezing of a collective atomic spin, 
Phys. Rev. Lett. {\bf 104}, 073602 (2011).

\bibitem{GrossNATURE2010}
C. Gross, T. Zibold, E. Nicklas, J. Est\`eve, and M. K. Oberthaler, 
Nonlinear atom interferometer surpasses classical precision limit, 
Nature {\bf 464}, 1165 (2010).

\bibitem{RiedelNATURE2010}
M. F. Riedel, P. B\"ohi, Y. Li, T. W. H\"ansch, A. Sinatra, and P. Treutlein (2010), 
Atom-chip-based generation of entanglement for quantum metrology, 
Nature {\bf 464}, 1170 (2010).

\bibitem{BohnetSCIENCE2016}
J. G. Bohnet, B. C. Sawyer, J. W. Britton, M. L. Wall, A. M. Rey, M. Foss-Feig, and J. J. Bollinger, 
Quantum spin dynamics and entanglement generation with hundreds of trapped ions, 
Science {\bf 352}, 1297 (2016).

\bibitem{StrobelSCIENCE2014}
H. Strobel, W. Muessel, D. Linnemann, T. Zibold, D. B. Hume, L. Pezz\`e, A. Smerzi, and M. K. Oberthaler, 
Fisher information and entanglement of non-Gaussian spin states, 
Science {\bf 345}, 424 (2014).

\bibitem{CiracPRA1998}
J. I. Cirac, M. Lewenstein, K. M\o lmer, and P. Zoller, 
Quantum superposition states of Bose-Einstein condensates, 
Phys. Rev. A {\bf 57}, 1208 (1998).

\bibitem{TrenkwalderNATPHYS2016}
A. Trenkwalder, G. Spagnolli, G. Semeghini, S. Coop, M. Landini, P. Castilho, L. Pezz\`e, G. Modugno, 
M. Inguscio, A. Smerzi, and M. Fattori, 
Quantum phase transitions with parity-symmetry breaking and hysteresis, 
Nat. Phys. {\bf 12}, 826 (2016).

\bibitem{KrusePRL2016}
I. Kruse, K. Lange, J. Peise, B. L\"ucke, L. Pezz\`e, J. Arlt, W. Ertmer, C. Lisdat, L. Santos, A. Smerzi, and C. Klempt, 
Improvement of an atomic clock using squeezed vacuum, 
Phys. Rev. Lett. {\bf 117}, 143004 (2016).

\bibitem{GabbrielliPRL2015}
M. Gabbrielli, L. Pezz\`e, and A. Smerzi, 
Spin-mixing interferometry with Bose-Einstein condensates, 
Phys. Rev. Lett. {\bf 115}, 163002 (2015).

\bibitem{WuPRA2016}
L.-N. Wu and L. You, 
Using the ground state of an anti- ferromagnetic spin-1 atomic condensate for Heisenberg- limited metrology, 
Phys. Rev. A 93, 033608 (2016).

\bibitem{SzigetiPRL2017}
S. S. Szigeti, R. J. Lewis-Swan, and S. A. Haine, 
Pumped- Up SU(1,1) Interferometry, 
Phys. Rev. Lett. {\bf 118}, 150401 (2017).

\bibitem{KajtochPREPRINT}
D. Kajtoch, K. Pawlowski, and E. Witkowska, 
Metrologically useful states of spin-1 Bose condensates with macroscopic magnetization, 
arXiv:1704.00628.

\bibitem{ZhangPRL2013}
Z. Zhang, and L.-M. Duan, 
Generation of massive entanglement through an adiabatic quantum phase transition in a spinor condensate, 
Phys. Rev. Lett. {\bf 111}, 180401 (2013).

\bibitem{HoangPNAS2016}
T. M. Hoang, H. M. Bharath, M. J. Boguslawski, M. Anquez, B. A. Robbins, and M. S. Chapman, 
Adiabatic quenches and characterization of amplitude excitations in a continuous quantum phase transition,
PNAS {\bf 113}, 9475 (2016).

\bibitem{LuoSCIENCE2017}
X-Y Luo, Y.-Q. Zou, L.-N. Wu, Q. Liu, M.-F. Han, M. K. Tey, and L. You, 
Deterministic entanglement generation from driving through quantum phase transitions,
Science {\bf 355}, 620 (2017).

\bibitem{LawPRL1998}
C. K. Law, H. Pu, and N. P. Bigelow, 
Quantum spins mixing in spinor Bose-Einstein condensates, 
Phys. Rev. Lett. {\bf 81}, 5257 (1998).

\bibitem{Stamper-KurnRMP2013}
D. M. Stamper-Kurn, and M. Ueda, 
Spinor Bose gases: Symmetries, magnetism, and quantum dynamics, 
Rev. Mod. Phys. {\bf 85}, 1191 (2013).

\bibitem{nota}
We have $\lambda = \frac{4 \pi \hbar^2 (a_2 - a_0)}{m} 
\int d^2 {\bf r} |\phi({\bf r})|^4$, where $\phi({\bf r})$, 
with $\int d^2 {\bf r} |\phi({\bf r})|^2=1$,  is the condensate (Gross-Piteaevskii)
wave function, $m$ the atomic mass and $a_G$ the scattering lengths for s-wave collisions in the $G = 0,2$ allowed channels.

\bibitem{nota1}
The ground state at $q=0$ is
$\ket{\psi_{\rm gs}^{q=0}} = \sqrt{\tfrac{2^N (N!)^3}{(2N)!}}\sum_{k=0}^{N/2}\tfrac{1}{2^kk!\sqrt{(N-2k)!}}|k\rangle$, 
see P. Feldmann et al. (in preparation).

\bibitem{HollandPRL1993}
M. J. Holland, and K. Burnett, 
Interferometric detection of optical phase shifts at the Heisenberg limit, 
Phys. Rev. Lett. {\bf 71}, 1355 (1993).

\bibitem{PezzeVarenna}
L. Pezz\`e and A. Smerzi, in {\it Atom Interferometry},
Proceedings of the International School of Physics ``Enrico Fermi'', Vol. 188, 
edited by G. M. Tino and M. A. Kasevich (IOS Press, Amsterdam, 2014) pp. 691-741;
arXiv:1411.5164.

\bibitem{LuckeSCIENCE2011}
B. L\"ucke, M. Scherer, J. Kruse, L. Pezz\`e, F. Deuretzbacher, P. Hyllus, O. Topic, J. Peise, W. Ertmer, 
J. Arlt, L. Santos, A. Smerzi, and C. Klempt, 
Twin matter waves for interferometry beyond the classical limit, 
Science {\bf 334}, 773 (2011).

\bibitem{HyllusPRA2012}
P. Hyllus, W. Laskowski, R. Krischek, C. Schwemmer, W. Wieczorek, H. Weinfurter, L. Pezz\`e, and A. Smerzi,
Fisher information and multiparticle entanglement, 
Phys. Rev. A {\bf 85}, 022321 (2012); 
G. T\'oth, 
Multipartite entanglement and high-precision metrology, 
Phys. Rev. A {\bf 85}, 022322 (2012).

\bibitem{PeiseNATCOMM2015}
J. Peise, I. Kruse, K. Lange, B. L\"ucke, L. Pezz\`e, J. Arlt, W. Ertmer, K. Hammerer, L. Santos, A. Smerzi, and C. Klempt, 
Satisfying the Einstein-Podolsky- Rosen criterion with massive particles, 
Nat. Comm. {\bf 6}, 1038 (2015).

\bibitem{GrossNATURE2011}
C. Gross, H. Strobel, E. Nicklas, T. Zibold, N. Bar-Gill, G. Kurizki, and M.K. Oberthaler, 
Atomic homodyne detection of continuous-variable entangled twin-atom states, 
Nature {\bf 480}, 219 (2011).

\bibitem{nota2}
As the QFI for pure state is given by the variance of an operator, MSS, 
which characterized by a large variance, have a large QFI.  
For instance a NOON state $(\ket{N,0} + \ket{0,N})/\sqrt{2}$ 
has a QFI at the Heisenberg limit, $F_Q = N^2$, while 
the incoherent mixture $(\ket{N,0}\bra{N,0} + \ket{0,N}\bra{0,N})/2$ has a QFI at the shot noise level, $F_Q=N$.

%
%




\end{thebibliography}

\end{document}